# SCMA for Open-Loop Joint Transmission CoMP

Usa Vilaipornsawai, Hosein Nikopour, Alireza Bayesteh, and Jianglie Ma
Huawei Technologies Canada Co., LTD. Ottawa, Ontario, Canada
{usa.vilaipornsawai, hosein.nikopour, alireza.bayesteh, jianglei.ma}@huawei.com

*Abstract*—Sparse Code Multiple Access (SCMA), a non-orthogonal multiple access scheme, has been introduced as a key 5G technology to improve spectral efficiency. In this work, we propose SCMA to enable open-loop coordinated multipoint (CoMP) joint transmission (JT). The scheme combines CoMP techniques with multi-user SCMA (MU-SCMA) in downlink. This scheme provides open-loop user multiplexing and JT in power and code domains, with robustness to mobility and low overhead of channel state information (CSI) acquisition. The combined scheme is called MU-SCMA-CoMP, in which SCMA layers and transmit power of multiple transmit points (TPs) are shared among multiple users while a user may receive multiple SCMA layers from multiple TPs within a CoMP cluster. The benefits of the proposed scheme includes: i) drastic overhead reduction of CSI acquisition, ii) significant increase in throughput and coverage, and iii) robustness to channel aging. Various algorithms of MU-SCMA-CoMP are presented, including the detection strategy, power sharing optimization, and scheduling. System level evaluation shows that the proposed schemes provide significant throughput and coverage gains over OFDMA for both pedestrian and vehicular users.

## I. INTRODUCTION

With high capacity demand in 5G wireless networks [1], the trend to deploy a large number of small cells, known as ultra-dense network (UDN), is envisaged [2]. However, such scenario posts challenges such as severe inter-cell interferences and mobility management, to name a few. Coordinated multipoint (CoMP), also known as network multiple-input multiple-output (MIMO) [3], [4], where multiple transmit points (TPs) are coordinated, is a key technology to mitigate such interferences. User multiplexing, where multiple users share the same resources, is another crucial technique to increase system capacity. Multi-user MIMO (MU-MIMO) based on beam forming can provide significant gain on average cell throughput [5], [6]. Moreover, the combination of MU-MIMO with coherent joint transmission (JT) CoMP provides further coverage gain [4]. Despite the promising benefits, both MU-MIMO and JT CoMP are in nature closed-loop schemes and hence are only applicable to low mobility scenarios due to their sensitivity to channel aging. Even for low mobility users, the sensitivity to quantization error of channel state information (CSI) and the cost of CSI acquisition[1] limit their practical gain. The problem becomes even more severe for a large cluster size as in UDN where a user is seen by several TPs. Therefore, an open-loop user multiplexing and JT CoMP scheme is desired with robustness to mobility, channel aging and low overhead of CSI acquisition.

Sparse code multiple access (SCMA) is introduced in [7] as a non-orthogonal multiple access scheme. Sparse codewords of multiple layers or users are overlaid in code and power domains and carried over shared OFDMA time-frequency resources. The system is overloaded if the number of overlaid layers is more than the length of multiplexed codewords. Overloading is achievable with moderate complexity of detection thanks to the sparseness of SCMA codewords. Inspired from low-density parity-check code (LDPC), message passing algorithm (MPA) is run over a sparse factor graph of SCMA to detect transmitted codewords with a near-optimal quality of detection [8]. In [9], downlink multi-user SCMA (MU-SCMA) is proposed to enable open-loop user multiplexing with robustness to mobility and low CSI feedback requirement. With a limited need for channel knowledge, a TP can pair users together while the transmit downlink power is properly shared among multiplexed layers in code domain. This technique relaxes the CSI feedback burden, and is robust to channel aging as compared to MU-MIMO technique, which is a closed-loop system.

In this work, we extend single-cell MU-SCMA to an open-loop JT CoMP solution so-called MU-SCMA-CoMP. SCMA layers and transmit power of TPs are shared among multiple users within a CoMP cluster. The transmit layer of these serving TPs are all multiplexed in SCMA code domain. A user may receive multiple SCMA layers from neighboring TPs and uses MPA to separate these layers and extract its intended data streams. Open-loop JT CoMP transmit diversity scheme is proposed in [10] based on space frequency block coding (SFBC) and cell-specific cyclic delay diversity (CDD) techniques, where different delays are applied to different cells. In [11], the extension of the scheme in [10] is designed to support demodulation reference signal (DM-RS) used in LTE-A system. Moreover, open-loop CoMP spatial multiplexing (SM) technique is considered in [11]. Although the schemes in [10], [11] are open-loop CoMP, they support only single-user CoMP (SU-CoMP) rather than multi-user CoMP proposed in this work.

Fig. 1 illustrates two application scenarios for open-loop SCMA CoMP for moving networks and UDN scenarios. Besides the throughput and coverage improvement of vehicular users, the proposed scheme can also facilitate soft handover for fast moving users as well as UDN in which frequent handover is a technical challenge. User-centric CoMP can be enabled in UDN via SCMA layer allocation across multiple TPs.

In this paper, various algorithms of MU-SCMA-CoMP are proposed for user pairing, layer allocation, power sharing and scheduling. System-level evaluation is performed to demonstrate the advantage of the proposed schemes in terms of cell throughput and coverage gains over baseline OFDMA system for both pedestrian and vehicular users.

---
[1] CSI acquisition can be either in the form of CSI feedback in FDD or channel sounding in TDD systems [12].

This paper is organized as follows. Section II presents SCMA system model, and its MIMO and linear-equivalent system. Section III discusses three MU-SCMA-CoMP schemes, each with detailed discussion on the detection strategy, user pairing algorithm and power sharing optimization. Section IV presents the scheduling algorithm. Simulation setup and results are shown in Section V, followed by conclusion in Section VI.

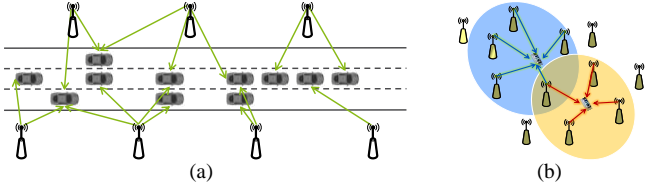

Fig. 1. (a) MU-SCMA-CoMP and inter-TP SCMA layer sharing even for vehicular users, (b) MU-SCMA-CoMP for UDN to enable user-centric clustering and soft handover.

## II. SYSTEM MODEL

In this paper, a network consisting of TPs and users are considered in which TPs transmit data to users in downlink. Users can be served by multiple TPs as in JT CoMP. Unlike the conventional CoMP schemes, open-loop transmission is assumed meaning that TPs only have channel quality indicator (CQI) of the users and not their corresponding CSI for example in terms of precoding matrix indicator (PMI) [12]-Table 7.2.1-1, which significantly reduces the feedback or sounding overhead. TPs are assumed to have single antenna while users can have multiple antennas. This assumption is made for simplicity of analysis. However, the results of this paper can easily be generalized to the case of multi-antenna TPs.

In order to support multi-user open-loop transmission in CoMP, TPs transmit their data using an SCMA encoder to multiple users (MU-SCMA-CoMP). As a codebook-based scheme, an SCMA encoder maps $\log_2(M)$ information bits directly to a $K$-dimensional complex codebook of size $M$. The $K$-dimensional complex codewords of a codebook are sparse vectors with $N < K$ non-zero elements. Sparsity helps to reduce the detection complexity by using MPA detector. In a single-TP SCMA system with $U$ users, let user $u, u = 1, 2, \ldots, U$ use the assigned $J_u$ layers. Denote $\mathbf{x}_{ju}$ of size $K \times 1$ as the $j$th SCMA codeword of user $u$, assuming $\|\mathbf{x}_{ju}\|^2 = K$, $p_u$ as the total transmit power per tone distributed equally over $J_u$ layers of user $u$. The total transmit power is $P = \sum_{u=1}^{U} p_u$ for all TPs. The $K \times 1$ received vector $\mathbf{y}_{ru_o}$ over an SCMA block of user $u_o$ at a received antenna $r$ can be expressed as,

$$\mathbf{y}_{ru_o} = \text{diag}(\mathbf{h}_{ru_o}) \sum_{u=1}^{U} \sqrt{\frac{p_u}{J_u}} \sum_{j=1}^{J_u} \mathbf{x}_{ju} + \mathbf{n}_{ru_o}, \quad (1)$$

where $\mathbf{h}_{ru_o}$ is the $K \times 1$ channel vector of $r$th receive antenna of user $u_o$ over $K$ OFDMA tones of an SCMA block, $\text{diag}(\cdot)$ is a diagonal operator and $\mathbf{n}_{ru_o}$ is an ambient AWGN noise vector. For simplicity, channel coefficients are assumed identical over $K$ tones of SCMA block, i.e. $\mathbf{h}_{ru_o} = h_{ru_o}\mathbf{1}$, where $\mathbf{1}$ is an all-one column vector.

### A. MIMO and linear-equivalent SCMA system for CoMP

Due to the non-linearity of SCMA modulation, it is not straightforward to model and obtain its capacity. Similar to [9], the MIMO and linear-equivalent of SCMA model is developed for multiple TPs to model the CoMP system. Consider a linear sparse sequence $\mathbf{x}_{ju} = q_{ju}\mathbf{s}_{ju}$, where $\mathbf{s}_{ju}$ with $\|\mathbf{s}_{ju}\|^2 = K$ is the $j$th signature of user $u$ and $q_{ju}$ is the corresponding QAM symbol with average unit power. In a $T$-TP system, let $\mathbf{y}_{u_o,t_o}$ denote the received signal at user $u_o$ associated with TP $t_o$

$$\mathbf{y}_{u_o,t_o} = \sum_{t=1}^{T} \sum_{u=1}^{U_t} \sqrt{\frac{p_{u,t}}{J_{u,t}}} \mathbf{H}_{u_o,t_o,u,t} \mathbf{q}_{u,t} + \mathbf{n}_{u_o,t_o}, \quad (2)$$

where $\mathbf{y}_{u_o,t_o} = (\mathbf{y}_{1u_o,t_o}^T, \ldots, \mathbf{y}_{ru_o,t_o}^T)^T$, $\mathbf{H}_{u_o,t_o,u,t} = \mathbf{h}_{u_o,t_o} \otimes \mathbf{S}_{u,t}$, with $\mathbf{h}_{u_o,t_o} = (h_{1u_o,t_o}, \ldots, h_{ru_o,t_o})^T$, $\mathbf{S}_{u,t} = (\mathbf{s}_{1u,t}, \ldots, \mathbf{s}_{Ju,t})$ being a signature matrix of $J_{u,t}$ signatures, $\mathbf{q}_{u,t} = (q_{1u_o,t_o}, \ldots, q_{Ju_o,t_o})^T$, is a vector of normalized QAM symbols, $\mathbf{n}_{u_o,t_o} = (\mathbf{n}_{1u_o,t_o}^T, \ldots, \mathbf{n}_{ru_o,t_o}^T)^T$, and $\otimes$ is Kronecker product operator.

## III. DOWNLINK JT CoMP USING MU-SCMA

In [9], downlink MU-SCMA is proposed to enable open-loop user multiplexing in a single-TP scenario. Similar to MU-SCMA, SCMA layers are shared among multiple users in downlink MU-SCMA-CoMP, and hence, transmit power needs to be shared among users. The difference in MU-SCMA-CoMP with the MU-SCMA is that a user can receive signal from multiple TPs. This means that the detection and user pairing strategies as well as power sharing optimization at a single TP affect the interference structure, which in turn affects the resulting rates of the paired users in the whole CoMP set. Therefore, in MU-SCMA-CoMP, all optimizations need to be done for all users and TPs as a whole, which makes the scheduling, user pairing and power allocation more challenging. In this paper, three options for MU-SCMA-CoMP are considered.

### A. Remote-pairing MU-SCMA-CoMP scheme

Consider the remote-pairing MU-SCMA-CoMP scheme shown in Fig. 2, where user $i$ is assumed to be a cell-edge user equipment (UE) associated with TP1, having low average signal to interference plus noise power ratio (SINR) so-called CoMP user, and user $j$ is a cell-centre user of TP2, having high SINR so-called good user. This scheme shows that TP2 can help the cell-edge user of TP1 by assigning power of $P_{i,TP2} = \alpha P$, while serving its own user with $P_{j,TP2} = (1-\alpha)P$, where $\alpha$ is a power sharing factor. With SCMA, it is possible to share layers, apart from sharing power among these users. This fact shows the MU aspect of the scheme, while the CoMP aspect is realized from the fact that user $i$ receives data from both TPs.

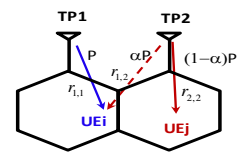

Fig. 2. Remote-pairing MU-SCMA-CoMP scheme.

In this work, we assume different data is transmitted from different TPs to the CoMP users and do not consider the transmission diversity mode in which the same data is transmitted from the TPs. This setup consists of multiple access channel (MAC) (i.e. from TP1 and TP2 to user $i$) and one broadcast channel (BC) (i.e. from TP2 to users $i$ and $j$). Hence, the results from information theory literature for both channels [13] are used to obtain rates for users in the system. Let $\mathbf{y}_{i,TP1}$ denote the received signal at user $i$ associated with TP1. Using (2), we have

$$\mathbf{y}_{i,TP1} = \sqrt{\frac{P}{J_{i,TP1}}}\mathbf{H}_{i,TP1,i,TP1}\mathbf{q}_{i,TP1} + \sqrt{\frac{\alpha P}{J_{i,T2}}}\mathbf{H}_{i,TP2,i,TP2}\mathbf{q}_{i,TP2} + \left(\sqrt{\frac{(1-\alpha)P}{J_{j,T2}}}\mathbf{H}_{i,TP2,j,TP2}\mathbf{q}_{j,TP2} + \mathbf{n}_{i,TP1}\right). \quad (3)$$

Assuming channel is known at the receiver, by treating signal for user $j$ sent from TP2 as interference, the achievable rate at user $i$ can be expressed as $\tilde{r}_i = \log_2 det(\mathbf{I} + \mathbf{R}^{-1}\mathbf{H}\mathbf{H}^H)$, where the equivalent channel is defined as $\mathbf{H} = \left[\sqrt{\frac{P}{J_{i,TP1}}}\mathbf{H}_{i,TP1,i,TP1} \quad \sqrt{\frac{\alpha P}{J_{i,T2}}}\mathbf{H}_{i,TP2,i,TP2}\right]$ and the covariance matrix of interference plus noise is given by $\mathbf{R} = N_i\mathbf{I} + \frac{(1-\alpha)P}{J_{j,TP2}}\mathbf{H}_{i,TP2,j,TP2}\mathbf{H}^H_{i,TP2,j,TP2}$, with $N_i$ being the ambient noise power at user $i$. Moreover, successive interference cancellation (SIC) detection can be used to achieve a corner point of the MAC capacity region. Considering the simplifications as in [9] to approximate the interference covariance matrix, we can express the data rates at user $i$ from TP1 and TP2, later used in power sharing optimization, as $\log_2 det\left(\mathbf{I} + \frac{\gamma^{CoMP}_{i,TP1}}{J_{i,TP1}(1+(1-\alpha)\gamma^{Co}_{i,TP2})}\mathbf{S}^H_{i,TP1}\mathbf{S}_{i,TP1}\right)$ and $\log_2 det\left(\mathbf{I} + \frac{\alpha\gamma^{CoMP}_{i,TP2}}{J_{i,TP2}(1+\gamma^{CoMP}_{i,TP1}+(1-\alpha)\gamma^{CoMP}_{i,TP2})}\mathbf{S}^H_{i,TP2}\mathbf{S}_{i,TP2}\right)$, where $\gamma^{CoMP}_{i,TPx} = \frac{\|\mathbf{h}_{i,TPx}\|^2 P}{N^{CoMP}_{i,TPx}}$ is the CoMP SINR, with $N^{CoMP}_{i,TPx}$ being the noise plus interference power from all cells outside the CoMP set of user $i$ associated with $TPx$. These rates are achievable if: i) the data for user $i$ from TP2 is first detected, treating signals for user $i$ from TP1 and user $j$ from TP2 as interference, and ii) the data for user $i$ from TP1 is detected, assuming only signal for user $j$ from TP2 as interference, while the signal for user $i$ from TP2 is assumed to be detected and cancelled out from the received signal. In case that $\alpha = 1$, this scenario falls back to SU-CoMP in which only one user is served by multiple TPs.

Let us consider the broadcast channel part of the system, i.e. TP2 transmits super-imposed signal (in the layer domain) to users $i$ and $j$, while treating signal transmitted for user $i$ from TP1 as interference. To simplify the analysis, the linear equivalent model is approximated by a degraded model by assuming the interference plus noise is white [9]. This means that the SIC detection can be used by ordering of the users in terms of increasing SINR. Assuming $\gamma^{CoMP}_{j,TP2} > \gamma^{CoMP}_{i,TP2}$, the SIC detection can be performed at user $j$. That is, the data for user $i$ from TP2 with lower $\gamma^{CoMP}_{i,TP2}$ is detected first, assuming the signal for this user from TP1 and that for user $j$ for TP2 as interference. In this degraded system with $\gamma^{CoMP}_{j,TP2} > \gamma^{CoMP}_{i,TP2}$, the data for user $i$ from TP2 can be detected at user $j$ and canceled out. Hence, the achievable rate at good user $j$ is approximated by $\tilde{r}_{j,TP2} \approx \log_2 det\left(\mathbf{I} + \frac{(1-\alpha)\gamma_{j,TP2}}{J_{j,TP2}}\mathbf{S}^H_{j,TP2}\mathbf{S}_{j,T2}\right)$, where $\gamma_{j,TP2} = \frac{\|\mathbf{h}_{j,TP2}\|^2 P}{N_{j,TP2}}$ is a non-CoMP SINR at user $j$, with $N_{j,TP2}$ defined as noise plus interference power from all TPs excepting the serving TP2 of user $j$. Note that we consider the non-CoMP SINR in the rate calculation of user $j$ since the signal for user $i$ from TP1 is an interference to this user.

For the special case of $\mathbf{S} = [1]$, all the obtained results apply to OFDM system which is degraded in nature. For an actual SCMA BC case, the assumption of $\gamma^{CoMP}_{j,TP2} > \gamma^{CoMP}_{i,TP2}$ does not guarantee that the user with lower SINR can be detected at higher SINR user. From the above discussion, we can summarize the detection strategy, power sharing optimization and greedy user pairing algorithm for this scheme as follows.

*1) Detection strategy*

At CoMP user, data from TP1 and TP2 can be jointly detected, assuming signal for good user from TP2 as interference. This means the data for this user from TP2 is first detected, assuming signal from TP1 and that for good user from TP2 as interference, then the data for this user from TP1 is detected, assuming only signal intended for good user from TP2 as interference. We can express the rates as

$$\tilde{r}_{i,TP1} = \log_2 det\left(\mathbf{I} + \frac{\gamma^{CoMP}_{i,TP1}}{J_{i,TP1}(1+(1-\alpha)\gamma^{CoMP}_{i,TP2})}\mathbf{S}^H_{i,TP1}\mathbf{S}_{i,TP1}\right),$$

$$\tilde{r}_{i,TP2} = \log_2 det\left(\mathbf{I} + \frac{\alpha\gamma^{CoMP}_{i,TP2}}{J_{i,TP2}(1+\gamma^{CoMP}_{i,TP1}+(1-\alpha)\gamma^{CoMP}_{i,TP2})}\mathbf{S}^H_{i,TP2}\mathbf{S}_{i,TP2}\right).$$

At good user, assuming a degraded system, the data for this user is detected considering only the signal for CoMP user from TP1 as interference. The rate can be expressed by

$$\tilde{r}_{j,TP2} = \log_2 det\left(\mathbf{I} + \frac{(1-\alpha)\gamma_{j,TP2}}{J_{j,TP2}}\mathbf{S}^H_{j,TP2}\mathbf{S}_{j,TP2}\right).$$

*2) Power sharing factor optimization*

The weighted sum rate (WSR) of these two users is used in the power sharing factor optimization based on maximum WSR criterion,

$$WSR_{i,j}(\alpha) = w_i(\tilde{r}_{i,TP1} + \tilde{r}_{i,TP2}) + w_j\tilde{r}_{j,TP2}.$$

The optimal $0 \leq \alpha^* \leq 1$ is the solution of $\frac{dWSR(\alpha)}{d\alpha} = 0$. Similar approach as in [9] can be taken to optimize $\alpha$. For OFDMA, the optimal $\alpha$ has a closed form expression as

$$\alpha^* = \frac{w_i\gamma^{CoMP}_{i,TP2}(1+\gamma_{j,TP2}) - w_j\gamma_{j,TP2}(1+\gamma^{CoMP}_{i,TP2})}{(w_i - w_j)\gamma^{CoMP}_{i,TP2}\gamma_{j,TP2}}, \quad (4)$$

where $w_i = \frac{1}{R_i}$, with $R_i$ being average rate of user $i$. For SCMA, there is no closed form for $\alpha^*$.

*3) Greedy user-pairing algorithm*

In first step user $i^*$ in TP1 is selected based on single-TP proportional fair (PF) metric $i^* = \arg\max_{u_i \in TP1} w_i r_i$. Then, for the given user $i^*$ user $j^*$ in TP2 is selected such that $WSR_{i^*,j^*}(\alpha^*)$ is maximized.

This greedy user-pairing algorithm can reduce a search space to only the number of users associated with cooperating TP, rather than all possible pairs of users between these two TPs. Notably, reciprocal procedure is also done at TP2, since it is possible that TP1 helps a CoMP user associated with TP2.

*B. Local-pairing MU-SCMA-CoMP scheme*

In local-pairing MU-SCMA-CoMP scheme shown in Fig. 3, two users are in the same cell. One is so-called the good user with high SINR, i.e. user $j$, another is so-called the CoMP user having low SINR (i.e. user $i$).

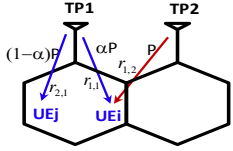

Fig. 3. Local-pairing MU-SCMA-CoMP scheme.

The detection strategy, power sharing optimization and greedy user pairing algorithm are summarized as follows.

*1) Detection strategy*

At CoMP user, similar to the remote-pairing scheme, we can express the rates as

$$\tilde{r}_{i,TP1} = \log_2 det\left(I + \frac{\alpha \gamma_{i,\,TP1}^{CoMP}}{J_{i,TP1}(1+(1-\alpha)\gamma_{i,\,TP1}^{CoMP}+\gamma_{i,\,TP2}^{CoMP})}S_{i,TP1}^H S_{i,TP1}\right),$$
$$\tilde{r}_{i,TP2} = \log_2 det\left(I + \frac{\gamma_{i,\,TP2}^{CoMP}}{J_{i,TP2}(1+(1-\alpha)\gamma_{i,\,TP1}^{CoMP})}S_{i,TP2}^H S_{i,TP2}\right).$$

At good user, assuming a degraded system, rate is expressed by

$$\tilde{r}_{j,TP1} = \log_2 det\left(I + \frac{(1-\alpha)\gamma_{j,TP1}}{J_{j,TP1}}S_{j,TP1}^H S_{j,TP1}\right).$$

*2) Power sharing factor optimization*

The power sharing factor optimization is based on maximum WSR criterion. For OFDMA, the optimal $\alpha$ is,

$$\alpha^* = \frac{w_i\gamma_{i,TP2}^{CoMP}(1+\gamma_{j,TP2}) - w_j\gamma_{j,TP2}(1+\gamma_{i,TP2}^{CoMP})}{(w_i-w_j)\gamma_{i,TP2}^{CoMP}\gamma_{j,TP2}}. \quad (5)$$

*3) Greedy user-pairing algorithm*

Similar to previous mode, user $i$ and $j$ both in TP1 are selected based on the greedy scheduling algorithm to maximize WSR.

*C. Dual-pairing MU-SCMA-CoMP scheme*

Fig. 4 shows the dual-pairing MU-SCMA-CoMP scheme, where user $j$ and user $k$ are cell-center users (or good users) with high SINRs, associated to TP1 and TP2 respectively, and user $i$ is a CoMP user with low SINR. This scheme consists of both remote and local pairing schemes. Moreover, it shows that both TPs serve a CoMP user, while serving their own users.

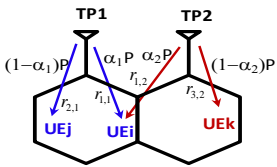

Fig. 4. Dual-pairing MU-SCMA-CoMP scheme.

*1) Detection strategy*

At CoMP user, data for this user from TP1 is first detected, assuming the other signals sent from TPs as interference. Then, the data for this user from TP2 is detected, assuming only signal intended for good users from TP1 and TP2 as interference. We can express the rates as

$$\tilde{r}_{i,TP1} = \log_2 det\left(I + \frac{\alpha_1 \gamma_{i,TP1}}{J_{i,TP1}(1+(1-\alpha_1)\gamma_{i,TP1})}S_{i,TP1}^H S_{i,TP1}\right),$$

$$\tilde{r}_{i,TP2} = \log_2 det\left(I + \frac{\alpha_2 \gamma_{i,\,TP2}^{CoMP}}{J_{i,TP2}(1+(1-\alpha_1)\gamma_{i,\,TP1}^{CoMP}+(1-\alpha_2)\gamma_{i,\,TP2}^{CoMP})}S_{i,TP2}^H S_{i,TP2}\right).$$

At good user $j$, assuming a degraded system, it is assumed that the data intended for the CoMP user from TP1 can be detected and completely canceled out. Hence, the rate can be expressed by

$$\tilde{r}_{j,TP1} = \log_2 det\left(I + \frac{(1-\alpha_1)\gamma_{j,TP1}}{J_{j,TP1}}S_{j,TP1}^H S_{j,TP1}\right)$$

At good user $k$, the data for good user $k$ is assumed detectable while signals from TP1 are interferences. The rate can be expressed by

$$\tilde{r}_{k,TP2} = \log_2 det\left(I + \frac{(1-\alpha_2)\gamma_{k,TP2}}{J_{k,TP2}}S_{k,TP2}^H S_{k,TP2}\right)$$

*2) Power sharing factor optimization*

The following WSR criterion is used in the power sharing factors optimization

$$WSR_{i,j,k}(\alpha_1,\alpha_2) = w_i(\tilde{r}_{i,TP1}+\tilde{r}_{i,TP2}) + w_j\tilde{r}_{j,TP1} + w_k\tilde{r}_{k,TP2}.$$

In an OFDMA system, for given users, $i$, $j$, $k$, power sharing is optimized by i) optimize $WSR_{i,j,k}(\alpha_1,\alpha_2)$ with respect to $\alpha_1$, to obtain $\alpha_1^* = f(\alpha_2)$, ii) substitute $\alpha_1 = \alpha_1^*$ in to $WSR_{i,j,k}(\alpha_1,\alpha_2)$ to obtain $WSR_{i,j,k}(\alpha_1^*,\alpha_2)$, iii) optimize $WSR_{i,j,k}(\alpha_1^*,\alpha_2)$ with respect to $\alpha_2$ to obtain $\alpha_2^*$, and iv) substitute $\alpha_2 = \alpha_2^*$ back to $\alpha_1^* = f(\alpha_2)$ to obtain $\alpha_1^*$.

For OFDMA, closed form expressions for $\alpha_1^*$ and $\alpha_2^*$ are

$$\alpha_1^* = 1 + \frac{w_i\gamma_{i,TP1}-w_j\gamma_{j,TP1}^{CoMP}+(1-\alpha_j^*)w_i\gamma_{j,TP1}^{CoMP}\gamma_{i,TP1}}{(w_i-w_j)\gamma_{j,TP1}^{CoMP}\gamma_{i,TP1}},$$
$$\alpha_2^* = 1 + \frac{(w_i-w_j)\gamma_{j,TP2}^{CoMP}\gamma_{i,TP2}+w_k\gamma_{k,j}(\gamma_{i,TP1}-\gamma_{j,TP1}^{CoMP})}{(w_i-w_j+w_k)\gamma_{j,TP2}^{CoMP}\gamma_{j,TP1}\gamma_{k,TP2}}. \quad (6)$$

*3) Greedy user-pairing algorithm*

The following steps are taken by the greedy scheduling algorithm: i) for a given pair of cooperating cells, e.g. TP1 and TP2, pick cell-center users, i.e. users $j$ and $k$ in Fig. 4, each based on the MU-SCMA user pairing, ii) search over all users (except user $j$) in TP1 to be a cell-edge user in the 3-user pairing, where power sharing factors are optimized for each paring based on (6), and iii) the 3-user pair with the maximum WSR is scheduled.

## IV. SCHEDULING ALGORITHM

At each scheduling interval, a set of users are scheduled in the cluster (all cells) such that among all the possible non-CoMP and CoMP modes, the one with the maximum WSR is selected. The detail procedure is explained in the following.

*A. Single-TP processing*

Each cell in the cluster performs a single-TP MU-SCMA scheduling as described in [9]. As the outcome of the scheduler either single or multiple users are served within every cell. The overall WSR of all cells in the cluster is calculated as the non-CoMP single-TP scheduling metric.

*B. Multi-TP processing*

For every possible CoMP set in the cluster, the best option among SU-CoMP, remote-, local-, and dual-pairing MU-SCMA-CoMP is dynamically selected in each scheduling interval. For the rest of the cells within the cluster, single-TP scheduling is performed as describe in previous sub-section. The overall WSR of the cluster is calculated by adding up the

WSR metrics of the entire cells of the cluster. It represents the metric of the multiple-TP scheduling.

The metrics of the single-TP and multiple-TP scheduling are compared and the best transmission mode is selected as the final outcome of the scheduling process to dynamically switch between non-CoMP and CoMP modes.

## V. SIMULATION SETUP AND RESULTS

System-level simulation for MU-SCMA-CoMP is performed on 7-cell cluster with hot cell (HC) in the middle. 70 users are dropped uniformly across the network with average 10 users per cell. For the sake of evaluation, only statistics of users in HC are collected. The channel is assumed to be frequency selective with coherence bandwidth of 1 MHz over 10 MHz system bandwidth centered at 2.6 GHz.

System-level performance results in terms of cell throughput and 5 percentile coverage rate of cell-edge users are presented in TABLE I. Both low and high user speeds of 3 km/h and 120 km/h are evaluated for 1×2 single-input multiple output (SIMO) scenario.

TABLE I. CELL THROUGHPUT AND COVERAGE RESULTS

| Case | Scenario | 3 km/h | | 120 km/h | |
|---|---|---|---|---|---|
| | | TPUT (Mbps) | Cov. (Kbps) | TPUT (Mbps) | Cov. (Kbps) |
| | **OFDMA** | | | | |
| 1 | OFDMA | 22.8 | 717.5 | 21.3 | 553.6 |
| 2 | OFDMA-CoMP | 23.2 | 810.7 | 20.9 | 651.9 |
| | **SCMA** | | | | |
| 3 | SCMA + SU-CoMP | 32.4 | 1021.9 | 28.9 | 805.8 |
| 4 | Case 3 + remote-pairing | 32.3 | 1045.8 | 28.8 | 822.9 |
| 5 | Case 4 + local pairing | 32.3 | 1062.5 | 28.9 | 847.9 |
| 6 | Case 5 + dual pairing (all possible modes) | 31.6 | 1219.4 | 28.7 | 977.5 |

Fig. 5 shows the throughput and coverage gains of SCMA over OFDMA baselines, i.e. OFDMA (Case1) and OFDMA-CoMP (Case2) for user speed of 120 km/h. The results show that the proposed MU-SCMA-CoMP schemes provide substantial gains, i.e. more than 35% (37%) cell throughput and 48% (26%) coverage gain over OFDMA (OFDMA-CoMP). The coverage gain of SCMA increases as more CoMP options are considered in the scheduling process.

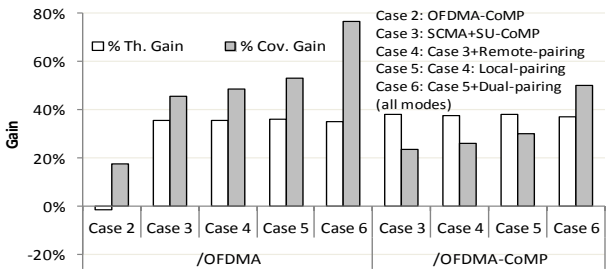

Fig. 5. Gains of MU-SCMA-CoMP over baselines: SIMO, 120 km/h

According to these results, MU-SCMA improves the average cell throughput while this user multiplexing technique in combination with the proposed CoMP modes can further improve the cell coverage. As the CoMP scenarios are extended over more users, the coverage performance is improved extensively at the expense of the negligible cell throughput drop. The slight drop of the cell average throughput is due to the fact that the resources and transmit power of cooperating cells are partially spent to improve the poor performance of cell-edge users rather than the multiplexing of multiple data streams for cell throughput enhancement. However, the benefit of CoMP for coverage enhancement is much more than its cell average throughout degradation. Note that a closed-loop scheme fails in this high speed scenario due to channel aging. This confirms the robustness of MU-SCMA-CoMP to user mobility.

## VI. CONCLUSION

In this paper, the extension of downlink MU-SCMA scheme for open-loop JT CoMP transmission is proposed and analyzed. Three open-loop MU-SCMA-CoMP schemes are developed for which the detection strategy, power sharing and user pairing algorithms are designed. Moreover, a scheduling algorithm is proposed to select the best transmission mode dynamically in each scheduling interval. System-level simulation results show the benefits of the proposed MU-SCMA-CoMP scheme in terms of average cell throughput and coverage over baseline OFDMA schemes even for high mobility scenario where the existing closed-loop solutions fail due to their practical limitations. Therefore, the proposed MU-SCMA-CoMP solutions pave the way to support better quality of service for various applications and scenarios in future 5G networks.